\documentstyle[aps]{revtex}
\begin{document}
\def\r{\hat\rho}
\def\s{\hat\sigma}
\title{Calibration of Local Chiralities and Time-Arrows Based on
Quantum Nonlocality}
\author{Lajos Di\'osi\thanks{E-mail: diosi@rmki.kfki.hu}\\
Research Institute for Particle and Nuclear Physics\\
H-1525 Budapest 114, POB 49, Hungary}
\maketitle
\begin{abstract}
Unlike classical physics, quantum mechnanics is sensitive to mistaken choice
of chiralities or time-arrows of local reference systems. Quantum 
correlations between distant electron spins, for instance, would reveal a
mistaken local chirality. Local polarization measurements and classical 
communication enable the distant partners to compare their local chiralities. 
Local time-arrows can be calibrated in a similar way.
\end{abstract}

Both the chirality and the time-arrow of a Cartesian reference system $xyzt$
are mathematically ambiguous. To settle the ambiguity we use phenomenological
conventions like, {\it e.g.}, the right-hand-rule for $xyz$ and the 
thermodynamic time-arrow for $t$. Since classical dynamics is fundamentally 
invariant for space- and time-reversals P and T, no absolute calibration of 
chirality and time-arrow exists in classical physics. We calibrate 
chiralities and time-arrows of separate local frames relative to each other. 
Elementary quantum mechanics is also invariant for P and T. It is, at the 
same time, more sensitive to local application of P or T than classical 
physics is. To utilize nonclassical correlations exhibited by the so-called 
entangled quantum states, I propose a certain quantum calibration of distant 
frames, which works even when its classical counterparts could not. At the 
heart of the method lies the mathematical discovery done by Stinespring 
\cite{Sti} as early as 1955. Many years after John Bell's famous work 
\cite{Bel}, it became a second cornerstone in quantum nonlocality studies 
\cite{Kra,Per,Hor,Ter}. Say, two distant quantum systems (Alice's and Bob's, 
resp.) form a composite quantum state $\r$. Then Alice applies, at least 
theoretically, a T or P transformation to her subsystem while Bob retains 
his:
\begin{equation}\label{rhoTP}
\r \rightarrow (T\otimes I)\r~~~or~~~\r \rightarrow (P\otimes I)\r~.
\end{equation}
Contrary to all classical evidences, the above local transformations may for
some $\r$'s {\it not\/} result in correct nonnegative density matrices. 
A necessary condition for this anomaly is that $T$ and $P$ must be 
anti-automorphisms \cite{Sti,Kra}. Indeed, time-reversal is always  
anti-unitary but space reversal $T$ is unitary. In this work, however,
we restrict the fundamental space-reflection for a particular one (while
notation $T$ is kept). We do not extend the operation $T$ {\it universally} 
for the (local) reference frame. To find an anti-unitary map, we have to 
restrict $T$ to certain local phenomena such as half-integer spins and we 
mean the reversal of the spin's Cartesian coordinates. This is equivalent 
with the {\it formal} reversal of the spin's reference frame and the 
reversal of the chirality of its axes $xyz$. This situation
occurs when the detected spin components are reversed e.g. by mistake.  
Space-reversal $T$, when understood in such specific and restricted sense 
for systems of half-integer spins, will be anti-unitary. 
In particular, for single electron spin vector $\s_n$ the map $T$ 
\begin{equation}\label{sigmaT}
\s_n\rightarrow -\s_n~~(n=1,2,3)
\end{equation}
requires an anti-unitary operator. 
Consequently, the corresponding composite quantum states may witness if 
Alice and Bob have mistakenly calibrated their local time-arrows or 
(half-integer-spin-frame's) chiralities \cite{Ebe}. Conditions regarding 
the class of `witness' states follow from results in Refs.~\cite{Per,Hor}.  

For concreteness, I consider the relative calibration of chiralities by Alice
and Bob respectively. Suppose they share a certain composite polarization 
state of electron pairs:
\begin{equation}\label{rho}
\r=\frac{1}{4}\bigl(1\otimes1
+\sum_n a_n \s_n\otimes1 + 1\otimes \sum_n b_n\s_n
+\sum_{n,m}c_{nm} \s_n\otimes\s_m\bigr)~,
\end{equation}
expanded in terms of the Pauli-matrices $\s$ and parametrized by the local 
polarization vectors $a,b$ and by their correlation tenzor $c$. For symmetry 
reasons, we assume that the electron pairs were prepared in a third reference
frame, say by Cecil, and the parameters $a,b,c$ refer to Cecil's frame. It is
most crucial that $a,b,c$ are constrained because the matrix $\r$ is 
nonnegative. Let the state (\ref{rho}) be a P-witness state. Now, we assume 
that Alice and Bob have {\it no prior information\/} on the state $\r$ and 
they do not know each others' or Cecil's reference frame either. Nonetheless,
Alice and Bob are measuring their local electron polarization vector 
$\s_1,\s_2,\s_3$ on a large statistics of pairs and, exchanging the 
measurement records via classical communication, they will in {\it bona 
fide} calculate a quantum state
\begin{equation}\label{rhobf}
\r^\prime=\frac{1}{4}\bigl(1\otimes1
+\sum_n a^\prime_n \s_n\otimes1 + 1\otimes \sum_n b^\prime_n\s_n
+\sum_{n,m}c^\prime_{nm} \s_n\otimes\s_m\bigr)
\end{equation}
on the basis of the measured polarization vectors $a^\prime,b^\prime$ and 
correlation tenzor $c^\prime$. If Alice and Bob's frames are rotated with 
respect to each other and/or to Cecil's then the bona-fide-state $\r^\prime$ 
will differ from the true state $\r$ (\ref{rho}). But Alice and Bob will not 
detect the discrepancy of states and frames. So far all is like in classical 
physics would be. However, a further discrepancy between the chiralities of 
their local frames can indeed be detected by Alice and Bob because the 
calculated bona-fide-state $\r^\prime$ will not exist at all. Alice and Bob 
will easily realize that the matrix (\ref{rhobf}) they have just calculated 
is indefinite.  

Let us see a simple example. Assume everyone has in reality the same 
Cartesian system $xyz$ but either Alice's or Bob's one is space-reflected. 
Neither Alice nor Bob know about this. They have received pairwise 
correlated electrons from Cecil. Each pair is in the totally anti-correlated 
singlet polarization state:
\begin{equation}\label{psi0}
\vert\psi\rangle=\frac{1}{\sqrt{2}}
\left(\vert+\rangle\otimes\vert-\rangle - 
      \vert-\rangle\otimes\vert+\rangle\right)
\end{equation}
but Cecil {\it does not inform\/} Alice and Bob about the state she prepared.
According to all classical wisdom Alice and Bob, relying on local 
measurements and mutual information communication, will never detect the 
mentioned discrepancy between their reference chiralities. Yet, quantum 
correlations make the detection possible. The density matrix of the pure 
state (\ref{psi0}) yields:
\begin{equation}\label{rho0}
\r\equiv\vert\psi\rangle\langle\psi\vert=
\frac{1}{4}\left(1\otimes1-\sum_n \s_n\otimes\s_m\right)~.
\end{equation}  
The average polarizations $a$ and $b$ vanish at both sites and the 
correlation is negative and isotropic: $c_{nm}=-\delta_{nm}$. When analysing
the records of their local measurements, Alice and Bob will obtain vanishing 
average polarizations $a^\prime,b^\prime$ and {\it positive\/} correlation 
$c^\prime_{nm}=\delta_{nm}$. The wrong sign is caused by the opposite 
reference chiralities of which Alice and Bob can not be aware. The 
bona-fide-state (\ref{rhobf}) would read:
\begin{equation}\label{rho0bf}
\r^\prime=\frac{1}{4}\left(1\otimes1+\sum_n \s_n\otimes\s_m\right)
\end{equation}
but it does not exist since its matrix has a negative eigenvalue. Indeed,
\begin{equation}\label{eigen}
\r^\prime\vert\psi\rangle=-\frac{1}{2}\vert\psi\rangle~,
\end{equation}
as it is easily seen from Eqs.~(\ref{psi0},\ref{rho0bf}). Hence, Alice and 
Bob realize immediately that the chiralities of their reference systems are 
incompatible. In the given context it is of course menaningless for Alice and
Bob to specify whose frame is mistaken. They will agree that one of them must
change the chirality of his/her local reference frame. Relative calibration 
of local chiralities is done.

Let us emphasize that Alice and Bob's test would be invariably conclusive if
Cecil submitted electron pairs in totally correlated nonproduct pure states
like, {\it e.g.}:
\begin{equation}\label{psi1}
\vert\psi\rangle=\frac{1}{\sqrt{2}}
\left(\vert+\rangle\otimes\vert+\rangle - 
      \vert-\rangle\otimes\vert-\rangle\right)~.
\end{equation}
The test is unconclusive if the state $\r$ is separable. Separable states
mean uncorrelated (product) states or statistical mixtures of such states. 
Mixing generates nothing but classical statistical correlations. Alice and 
Bob need nonseparable states exhibiting quantum-correlations as well. 

Let us extend the interpretation of the density matrix (\ref{rho}) for 
whatever two-by-two-state composite systems. Following Peres' prediction 
\cite{Per}, the Horodeckis' \cite{Hor} point out that all nonseparable 
quantum states in $2\times2$ dimensions witness the transpose of the local
density matrix owned {\it e.g.} by Alice. This purely mathematical operation 
is, in the usual representation of Pauli-matrices, equivalent to the 
sign-reversal of Alice's operator $\s_2$ and this is equivalent to the local 
time-reversal $T$. Therefore all nonseparable states are T-witnesses 
\cite{San}. The sign-reversal of $\s_2$ is, upto a unitary rotation, 
equivalent to space-reversal provided the system is a 
$\small{\frac{1}{2}}$-spin. Then, all nonseparable states are P-witnesses. 

We can shortly discuss the calibration of local time-arrows. As we said 
above, all nonseparable states of $2\times2$ dimensional composite systems 
are T-witnesses. Alice and Bob can use electrons as well as photons or atoms.
The state can still be written in the form (\ref{rho}), the Pauli-matrices
and the parameters will of course change their physical interpretations.
Otherwise, the relative calibration of Alice and Bob's local time-arrows is
completely analogous with the calibration of chiralities. Of course, the
postulated conditions are even more artificial. Few would beleive in the
reality of mistaken local time-arrows. Furthermore, Alice and Bob should not 
leave any timing information in their local measurement records. Yet, these 
hypothetical conditions are logically consistent. We can thus think of
time-reflection as simulated, {\it e.g.}, by the electronics operating 
typically between the measured quantum system and the final records.
 
One can not escape discussing the present proposals in the light of particle 
physics. Why, the steadfast belief in fundamental P and T invariances of 
quantum physics had been overthrown long ago. In 1957 experiments confirmed 
that neutron decays broke the left-right symmetry \cite{noT}. Later in 1964, 
the decay of neutral kaons proved, in an indirect way, to violate the 
time-reversal invariance \cite{noT}. Now Alice and Bob do in principle have 
a perfect test: by exchanging measurement records on local neutron and kaon 
decays they can reveal if their local reference systems happened to have 
different chiralities and/or different time-arrows. They do not need to share
entangled quantum states either. The tests, presented in my work, are not
substitutes of the tests based on fundamental symmetry violations particularly
because my tests can not detect a {\it universally} mistaken local chirality
\cite{Ebe}.  

Let us summarize the points of my work. In both classical and quantum
physics, relative calibration of local chiralities and time-arrows can be
based on chiralities and timings of local physical objects. For successful 
calibration, an accompanying information is needed because the distant
calibrating partners have to know which way the chiralities and timings 
of their local objects, respectively, are correlated between the different
sites. In the case of quantum objects, this information may be spared provided
the composite state of the objects is nonseparable, {\it i.e.}, it exhibits
quantum correlations. Quantum calibration needs less resources than its
classical counterpart and this is a remarkable result in spite of the
apparently artificial constraints postulated for the calibration. The present 
proposal confirms a conjecture that quantum nonlocality, a genuine 
physical feature of composite systems, is not `exploitable' directly in a
pure physical context. Advantage of quantum nonlocality manifests itself
rather in contexts including physics {\it and} information, cryptography 
\cite{Cry}, computation \cite{Sho}, or just games \cite{Gam}. In a very 
recent work \cite{Joz}, Jozsa {\it et al.} exploit quantum nonlocality's 
advantage in synchronizing local clocks. Calibration tasks seem, in certain
circumstances, to priviledge methods of quantum-correlations.    

I am indebted to Philippe Eberhard for his valuable remarks on the
former version of this paper. This work was supported by the Hungarian OTKA 
Grant 32640.

\end{document}